# Nature of low dimensional structural modulations and relative phase stability in $MoS_2$/$WS_2$-$ReS_2$ transition metal dichalcogenide alloys

R. Sahu,[1] U. Bhat,[1] N. M. Batra,[3] H. Sharona,[1] B. Vishal,[1] S. Sarkar,[2] S. Assa Aravindh,[3] S. C. Peter,[2] I. S. Roqan,[3] P. M. F. J. D. Costa,[3] and R. Datta[1,3,*]

[1]*International Centre for Materials Science, Chemistry and Physics of Materials Unit, Jawaharlal Nehru Centre for Advanced Scientific Research, Bangalore 560064, India.*
[2]*New Chemistry Unit, Jawaharlal Nehru Centre for Advanced Scientific Research, Bangalore 560064, India.*
[3]*Physical Science and Engineering Division, King Abdullah University of Science and Technology (KAUST), Thuwal, 23955-6900, Saudi Arabia.*

*Corresponding author E-mail: ranjan@jncasr.ac.in*

We report on the various types of Peierls like two dimensional structural modulations and relative phase stability of 2H and 1T poly-types in $MoS_2$-$ReS_2$ and $WS_2$-$ReS_2$ alloy system. Theoretical calculation predicts a polytype phase transition cross over at ~50 at.% of Mo and W in $ReS_2$ in both monolayer and bulk form, respectively. Experimentally, two different types of structural modulations at 50% and a modulation corresponding to trimerization at 75% alloy composition is observed for $MoS_2$-$ReS_2$ and only one type of modulation is observed at 50% $WS_2$-$ReS_2$ alloy system. The 50% alloy system is found to be a suitable monolithic candidate for metal semiconductor transition with minute external perturbation. $ReS_2$ is known to be in 2D Peierls distorted $1T_d$ structure and forms a chain like superstructure. Incorporation of Mo and W atoms in the $ReS_2$ lattice modifies the metal-metal hybridization between the cations and influences the structural modulation and electronic property of the system. The results offer yet another effective way to tune the electronic structure and poly-type phases of this class of materials other than intercalation, strain, and vertical stacking arrangement.

## I. Introduction

Atomically thin transition metal dichalcogenides (TMDs) is considered to be the next generation platform for future electronics and offers numerous novel device applications based on its unique excitons, spin, and valley properties [1-10]. The most explored members in the family i.e. $MoS_2$ and $WS_2$ posses direct band gap in the monolayer form but undergoes transition to undesirable indirect band gap material for number of layers two and more [5]. The stable crystal structure of $MoS_2$ and $WS_2$ is 2H (space group 194: P6$_3$/*mmc*) with a direct band gap of 1.88 and 1.9 eV, respectively for the monolayer. On the other hand for $ReS_2$, the band gap remains direct even in the bulk form due to weak interlayer van der Waals and electronic coupling [11]. The stable crystal structure of $ReS_2$ is 2D-Peierls distorted $1T_d$ (space group 2: $P\bar{1}$) where chains made of $Re_4$ clusters form the quasi two dimensional superstructure with a band gap of 1.55 eV for the monolayer. The metastable unmodulated 1T phase of $MoS_2$ is demanding due to its metallicity which is indispensable for carrier transport, injection and in the modulated $1T_d$ form shows excellent hydrogen evolution activity (HER) [12]. The most stable configuration in the modulated 1T form of $MoS_2$ and $WS_2$ is $2a \times a$ superstructure as in the case of $ReS_2$ [13-15]. These superstructures reported to have band gap of 0.1-0.2 eV and 0.14 eV, for $MoS_2$ and $WS_2$, respectively [14, 15]. Throughout the text '$1T_d$' is used to represent the chains like superstructure in all cases.

As already mentioned above, the 1T and modulated $1T_d$ poly-type forms of both $MoS_2$ and $WS_2$ are metastable. However, due to their useful properties various attempts have been made to stabilize these poly-type structural forms [16-19]. Most of the reports are based on intercalation method using alkali metals e.g., Li, and K where structural transitions between various poly-types take place during loading of alkali metals at the interlayer spacing with the end application as rechargeable batteries. Among various reports, Li and alkali metal intercalation was reported to stabilize the $1T_d$ form of both $MoS_2$ and $WS_2$ through hybridization between cations forming quasi 2D chain. Li intercalation of $MoS_2$ powders in water formed $2a \times 2a$ superstructure as confirmed by X-ray diffraction [20]. On the other hand, K intercalation stabilizes two different form of modulated structure of $MoS_2$ depending on the K concentration i.e. trimerization of Mo with superstructure $(a\sqrt{3} \times a\sqrt{3})$ ($x \approx 0.3$) and tetramerization of Mo with superstructure $2a \times 2a$ ($x \leq 0.3$) [21]. The presence of two different superstructures in this case was confirmed by scanning tunneling microscopy (STM) technique.

Relative stability of various types of superstructure modulation in Li intercalated $MoS_2$ i.e. $LiMoS_2$ was previously investigated by first principle calculations [22]. It was found that the $(2a \times 2a)$ superstructure structure formed by tetramerization of cations was more stable (by 0.5 eV) compared to trimerized $(a\sqrt{3} \times a\sqrt{3})$ structural modulation. This opened a band gap of about 1 eV. However, later studies based on both experimentation and theory revealed that it is $2a \times a$ type periodicity forming chains like structure the most stable modulated form of $MoS_2$ [13]. It was also pointed out that a charge density wave (CDW) phenomenon was at the origin of such structural modulation which was associated with a partially nested Fermi surface and the stabilization is localized around the Fermi level. The destabilization due to strain occurs mainly in the $3p$ S part which is compensated by electronic stabilization in the $3d$ Mo part. Though energetically least stable, however, the trimerized modulated structure was predicted to be the thinnest ferroelectric material [23].

There is a lot of interest in the community to understand and control various structural poly-types of monolayer TMDs by different means for their efficient utilization in energy, optoelectronics, and novel devices apart from immense interest in fundamental science [3-8, 24-27]. Among recent reports, intermediate steps of structural transition from trigonal prismatic 2H to octahedral 1T phase have been studied by electron microscopy [18]. Three different types of distortions or sub-superstructures are observed by HRTEM as transient states involving a shear mechanism. These are (2×1), (2×2) and ($\sqrt{3} \times 1$) and the combinations between them. The $d$ electron system for various superstructures changes as follows, $d^2$ for 2×1 (zigzag clustering), $d^{3/4}$ for ($\sqrt{3} \times 1$) (ribbon chain clustering) and $d^3$ for 2×2 (diamond chain clustering). In another report, Li intercalated $MoS_2$ shows the presence of bi-phase structure i.e. both 2H and combination of 1T and $1T_d$ (2×1) [1]. Different structural phases form coherent interface at the boundary. The structure changes from $1T_d$ to 1T by $e^-$ beam exposure. The transition from $1T_d$ to 2H is a multi step process and goes via different metastable structures. In the case of $WS_2$, 1T is found to be more stable than $MoS_2$. After annealing at different temperature the structure does not transform to 2H phase completely and defects are believed to stabilize the residues of 1T phase.

Formation of metastable tetramer diamond unit along with zigzag chains with a new orientation formed by $e^-$ beam irradiation of zigzag 1T phase of Li intercalated $WS_2$ has been reported [14]. In the 1T zigzag starting phase both tetramer and triangular clusters were present. The presence of triangular cluster in K intercalated $2H-MoS_2$ has already been predicted theoretically [16] and experimentally confirmed [28]. Theoretically the energy difference between the two meta-stable structures is found to be 0.1 eV per formula unit. The stabilization of the local tetramer in the present case is due to electronic charging effect. It was argued that the observed distortion is not a charge density wave (CDW) phenomena but rather Jahn Teller (JT) type introduced by a weakening of the W-S bonds leading to electronic instability. However, ReS$_2$ is not a Jahn Teller type as argued before because the gap opens in the *d* bands at the Fermi level as confirmed by modern first principle calculations [10, 29, 30].

Thus, in the present report, stabilization of 1T$_d$ structural form of both MoS$_2$-ReS$_2$ and WS$_2$-ReS$_2$ alloys has been explored by both theoretically and experimentally. Theoretically, a structural cross over is observed at approximately 50% alloy composition in both monolayer and bulk form. In this context, the stability of metallic 1T phase of MoS$_2$ upon 25% Sn substitution was predicted theoretically [31]. There are some reports where Re incorporation in MoS$_2$ nanotube, fullerene, and nanoparticles forms reported to show beneficial effects on HER activity [32-37]. Taking theory as a guide experimentally only 50% alloy composition is synthesized in both the cases. However, for MoS$_2$-ReS$_2$ alloy two different compositions i.e. 50 and 75% and for WS$_2$-ReS$_2$ case mostly homogeneous 50% alloy composition was obtained after the synthesis procedure. In the case of MoS$_2$-ReS$_2$ alloy, two different types of modulations i.e. tetramer (type I) and trapezoid like modulations (type II) or two different modulations between two different cations for 50% and a trimer modulation is observed for 75% alloy composition. The observation of stable trimerized structural modulation is important as this could be useful as a thinnest platform for ferroelectric devices [23]. On the other hand in WS$_2$-ReS$_2$, the alloy formation is homogeneous and only one type i.e. tetramer modulation (type I) is observed along with some areas with different modulation vectors probably due to slightly varying composition in these areas. The energy difference between the metallic and semiconducting ground states with a slightly differing structural parameters is found to be very small; 90 and 40 meV for the 50% $MoS_2$-$ReS_2$ and $WS_2$-$ReS_2$ alloys, respectively. This could be an excellent monolithic candidate for the metal-semiconductor transition with minute external perturbation suitable for application as a nanoscale switching device. Overall, the results show that the alloy can be formed between these important materials system with a differing stable structural form of the terminal compounds i.e. 2H and $1T_d$ and the various structural modulations observed offers yet another way to controlling the electronic structure of this system other than intercalation, strain, and vertical stacking arrangement. While high resolution electron microscopy revealed the details of such modulated structures, the density functional theory based calculations throw light on the implications on the electronic properties of such alloy system.

## II. Experimental Procedures

$Re_{0.5}X_{0.5}S_2$ (X = Mo and W) alloys were synthesized following a similar synthetic procedure reported earlier for $ReS_2$ [30]. In short, rhenium (wire, 1.0 mm diameter, 99.97%, Alfa Aesar), molybdenum/tungsten (powder, 99.9%, Sigma Aldrich / powder, 99.8%, Sigma Aldrich) and sulfur (powder, -325 mesh, 99.5%, Alfa Aesar) were taken in the stoichiometric ratio (1:1:4) and sealed in an evacuated ($10^{-5}$ mbar pressure) quartz tube. The tube was then heated to 200 °C at the rate of 10 °C/h followed by annealing at that temperature for 2 h in order to avoid any possible explosion due to the high vapor pressure of sulfur. Next, the temperature was raised to 900 °C at a rate of 20 °C/h and annealed for 120 h after which the furnace was shut down and the sample was allowed to cool to room temperature naturally. The final product was fine black powder.

All the high resolution phase contrast transmission electron microscopy (HRTEM, in ICMS, India) and Z-contrast high angle annular dark field imaging (HAADF, in KAUST, Saudi Arabia) were performed in a FEI TITAN aberration corrected 80-300 keV TEM. Powder alloy samples were sonicated for 40 minutes to exfoliate monolayer materials for TEM imaging and energy dispersive spectroscopy (EDS) analysis.

### III. Theoretical Calculations

The Electronic structure calculations were performed using density functional theory (DFT) as implemented in Wien2k code [38]. Wien2k uses linearized augmented plane waves as the basis and considers all electrons into the calculation. We have used the following structures for the present calculation: monolayer and bulk form of 2H & 1T $MoS_2$, $WS_2$, $ReS_2$, various alloys between them and also considering structural modulations for $ReS_2$ and some of the intermediate alloy composition. (The schematic of calculated structures are shown in various figures while describing the results and also some in the supplementary document). For the construction of bulk (1T/2H) and monolayer (1T/2H) $Re_xMo/W_{(1-x)}S_2$ alloy structure, various sizes of supercell depending on Re concentrations was constructed e.g., (i) 4×4×1 supercell for 6.25% Re, (ii) 4×2×1 for 12.5% Re, (iii) 2×2×1 for 25% and 75% Re, and (iv) 2×1×1 for 50% Re. A Γ centered 30×30×1 and 30×30×18 k point mesh was used for single formula unit cell of monolayer and two formula unit cell of bulk. The number of the k-mesh points was reduced proportionally for the larger supercell.

For bulk $MoS_2$ and $WS_2$ calculation, a unit cell consisting two formula unit was considered with Bernal stacking along the *c*-direction. DFTD3 was introduced for dispersion correction which considers van der Waals interaction between the layers [39]. In case of monolayer calculation, 16 Å vacuum was inserted between the layers in order to prevent interlayer coupling. The optimization of lattice parameters and atomic position for the $MoS_2$, $WS_2$ and distorted $ReS_2$ model were performed using the generalized gradient approximation (GGA) exchange-correlation functional of Perdew, Burke, and Enzerhof (PBE) [40]. The criteria of convergence for force, energy and electronic charge were set below 1 mRy/au, 0.0001 Ry and 0.001 e, respectively. The relaxed lattice parameters of $MoS_2$, $WS_2$, modulated $ReS_2$, and all other alloy structures considered for the present calculations are shown in Table 2 (Supplementary). The muffin tin radii and RMT were chosen in such a way that they do not overlap. The $K_{max}$ was set at 7.00/RMT.

## IV. Results and Discussion

We begin with the experimental structural description and associated superstructures observed by high resolution transmission electron microscopy in both $MoS_2$-$ReS_2$ and $WS_2$-$ReS_2$ alloy systems. As mentioned earlier, the most stable structure of $ReS_2$ is $1T_d$ (space group 2: P-1) and for $MoS_2$ and $WS_2$ is 2H (space group 194: P6$_3$/*mmc*). $ReS_2$ is known to be in a 2D Peierls distorted $1T_d$ structure where four Re atoms hybridize to form $Re_4$ clusters. Incorporation of Mo and W atoms in the $ReS_2$ lattice modifies the metal-metal hybridization between the cations and influences the structural modulation and electronic property of the system. Figure 1 (a) is the typical $1T_d$ $ReS_2$ structure where four Re atoms together hybridize to form tetramer $Re_4$ cluster forming quasi two dimensional chains like superstructure. The calculated superlattice spots corresponding to 1T (one of the S hexagons is rotated by 30° with respect to other Re and S hexagons) and periodic modulations/superstructure ($2a \times a$) are indicated with two different color circles. The lattice parameter of monolayer $ReS_2$ is $a$ = 6.4308 Å, $b$ = 6.4912 Å and $\gamma$ = 119.03°. The typical stable 2H and distorted 1T structure of $MoS_2$ and $WS_2$ are shown in the Fig. 1(b) & (c). Kindly note, that the length of the edges and the angle between them of the modulated structural units can be different along different directions for the same superstructure periodicity depending on the alloy composition. Therefore, in order to distinguish such modulations between different alloy systems we take help from the geometry by drawing a rhombus with four cations at the corners as shown in Fig. 1(d). Two different directions are defined as $a_1$ and $a_2$, with an obtuse angle between them as $\gamma$. Another two parameters defined as $a_1´$ and $a_2´$ along the original $a_1$ and $a_2$ directions, which will define the distance between the hybridized units along the two different directions. The supercell lattice parameters of the system are also defined in terms of repeat vectors $a$ and $b$ [Fig. 1(d)]. This is helpful to describe in details the nature of modulation vector and their strength along distinct directions, which in the literature either various nomenclature or dimensionality of modulations (1D or 2D) was used to define such structures. However, distortions can be quasi 2D with different lattice vectors and strength along different directions and therefore, we have used this definition to distinguish the various structural distortions observed in both the alloy systems. Additionally, the bonding lines between cation-cation represent hybridization and various types of hybridization possibilities are also indicated in the same figure. For both the alloys, all the structures are found to be the $1T_d$ poly-type form with different types of superstructures depending on the alloy composition. All the structural parameters obtained from theoretical and experimental observations are given in the supplementary document.

### A.    Alloy of $MoS_2$-$ReS_2$:

Three different types of structural modulations or super structures are observed in this alloy system depending on the composition. Out of three, two different modulations are observed for the same 50% and the third one with a trimerize type modulation for 75% $MoS_2$-$ReS_2$ alloy case. Percentage numbers in the alloys indicate the composition in terms of either Mo or W in the host $ReS_2$ lattice. There are areas where pathways of poly-type structural transition are observed due to gradual changes in Re concentrations and are also described.

(a) **Type I modulation of 50% $MoS_2$-$ReS_2$:**

Figure 2 (a) shows the modulation for the monolayer alloy. Alternate rows are filled by either Re and Mo atoms. This gives rise to the different modulations of lattice parameters along different directions. This is schematically shown in the inset. The Mo-Mo, Re-Re, and Re-Mo distances are 2.88, 2.80 and 2.68 Å, respectively. The acute angle $\gamma$ of the rhombus is 74°. $a_1'$ and $a_2'$ are 2.9, and 2.7 Å. The FFT (fast Fourier transformation) equivalent to the periodic diffraction pattern of the image is shown in Fig. 2(d). FFT also shows the signature of the $1T_d$ phase. The superlattice spots corresponding to $1T_d$ phase is strong along one of the directions which correspond to the parallel planes containing only Re and Mo atoms. The observed modulation in this case is similar to the $ReS_2$ but with different lattice vectors. The Z-contrast line scan profile identifying Re and Mo in a column are shown in the Fig. S2 (Supplementary).

(b) **Type II modulation of 50% $MoS_2$-$ReS_2$:**

For the same 50% alloy composition, a unique structural modulation is observed in some areas particularly near the surface regions and to best of our knowledge this type of structural modulation is not known in the literature [Fig. 2 (b)]. Two consecutive parallel lines are marked along which modulation of cations are different i.e. one short and another large distance periodic structural modulation. This can also be thought of in terms of two independent one dimensional Peierls distorted atomic chains or 2D type with doubling the edge of one side of the supercell unit. Considering all four atoms from the consecutive rows, a trapezoid like geometry can be identified or in other words the structural parameters can be described using two numbers of rhombuses and doubling the side along one direction (inset of Fig. 2(b)). The lattice parameters are shown in the same figure. FFT also reflects the corresponding symmetry [Fig. 2 (e)]. The distances between the atoms which are short and long are 2.42 Å and 2.69 Å, respectively. This trapezoid structure can be stabilized in both tetramer type of modulation in like $Re_4$ in $ReS_2$ (2D chain) and NN configurations (1D chain) depending on Re concentration as shown in Fig. S 3.I (supplementary).

**(c) Trimerized modulation of 75% $MoS_2$-$ReS_2$:**

In this case, a trimerized modulation is observed as shown in Fig. 2(c). The triangles are indicated in the figure along with the schematic (inset in Fig. 2(c)) and the length of the edges is equal to each other and is 3.2 Å. At the center of the large hexagon Re atom is occupying the site as marked with the blue circles. The structure can also be described by two parallel lines with one of them completely occupied by Mo atoms and another one is alternately filled by Mo and Re atoms. The distance between the Mo-Re atom is 2.8 Å. The $\gamma$ is 109° in this case. The FFT shows the signature of $1T_d$ phase along with structural modulation [Fig. 2 (f)]. The observation of stable trimer modulation is significant in the sense that such a structure was predicted to be the thinnest ferroelectric in the case of $MoS_2$ [23]. The present alloy offers a stable trimer system without intercalation route and might be amenable for a possible thinnest ferroelectric device fabrication. The Mo-Mo distance in the trimmer is ~ 3.2 Å. The Re-Mo or Re-Re bond distance in trimmer is ~ 2.8 Å to 2.9 Å as shown in the HRTEM images in the Fig. S4 (Supplementary).

In some areas pathways of phase transition between 2H to $1T_d$ structure is observed due to gradual changes in Re concentrations as shown by both HRTEM and HAADF images [Fig. 3 & 4]. This is an accidental finding and involves different intermediate phases depending on local Re distributions. Four distinct intermediate transition phases have been identified as shown in Fig. 3. The transition from 2H to $1T_d$ takes place via simultaneous rotation and gliding of one of the two S planes of the 2H crystal structure. As mentioned before, that Re-Re, Re-S bond length are shorter compared to Mo-Mo, Mo-S counterpart because of the smaller ionic radius of Re. The gradual decrease in Metal-S bond length due to introduction of Re atoms at Mo sites can be observed for the intermediate phases in going from 2H to 1Td and these are indicated and tabulated in the same figure. The intermediate phase IV is similar to the intermediate stripe phase reported earlier for the hydrogenated 2H $MoS_2$ surface [41]. As already mentioned, the presence of four different phases are due to different distribution of Re in the $MoS_2$ lattice which can clearly be observed in Z contrast image and the route from 2H to $1T_d$ transition is indicated by arrows [Fig. 3]. This image shows clearly the movement of S plane due to Re incorporation in the lattice and extent to which the plane will move depends on the local population of Re. Fig. 4 (b)-(e) are showing the atomically resolved images from areas with single Re substitution, two Re atoms forming a dimmer, three Re atoms forming a trimmer, four Re forming a tetramer, respectively. The various metal-metal and metal-S bond lengths are also mentioned. It is found that the smallest Re-S bond length in mono atomic doping is 1.39 Å ($Re_1$-$S_1$) and for dimmer is 1.42 Å ($Re_2$-$S_3$) and 1.65 Å ($Re_1$-$S_3$). The area where four Re atom forming tetramer the metal-metal distance is 2.7 Å and 2.9 Å which is similar to Re-Re distance in $ReS_2$. Example line profile identifying two different metals for various clustering are shown in the inset. A gradual S plane rotation (along with gliding of plane) is observed from 2H to 1T phase and associated modulation of structure due to hybridization between cations. A detailed schematic representation showing the correlation between the movements of S plane with Re content for various intermediate phases observed experimentally are given in Fig. 3 and this contain very rich structural variation in this 2D alloy system and worth investigating the electronic properties of such intermediate phases part of further work.

### B. Alloy of $WS_2$-$MoS_2$:

In this case, the alloy composition (i.e. 50%) is homogeneous in most of the regions probed and no variation in structural modulation is observed unlike $MoS_2$-$ReS_2$ alloy system except few areas where the deviation in modulation is observed probably due to slight variation in composition. This is due to good mixing between W and Re atoms as they are next to each other in the Periodic Table and have almost equal atomic size. This slight variation in the composition cannot easily be determined between W vs. Re atoms, unlike Mo vs. Re atoms where atomic number difference is high. We describe the detailed observation below;

#### (a) Type I modulation of 50% $WS_2$-$ReS_2$:

This is exactly similar to the 50% $ReS_2$-$MoS_2$ alloy system described earlier with alternating layer of W and Re atoms as indicated in Fig. 5 (a). This gives strong superlattice spots in the FFT image which corresponds to the parallel planes containing mono-type atoms [Fig. 5 (b)]. This is a quasi two dimensional system and the geometrical parameters are given in Table 3 in supplementary which are slightly different compared to the $MoS_2$ counterpart. The angle $\gamma$ is 80° between the $a_1$ and $a_2$ directions. The W-W, Re-Re, and Re-W distances are 2.78, 2.68 Å, and 2.43 Å, respectively. We do not observe type II modulation in this case like in $MoS_2$-$ReS_2$ alloy system.

#### (b) Variation in Type I modulation in $WS_2$-$ReS_2$:

In some areas variation in the lattice parameters of superstructure modulation is observed [Fig. 5 (c)]. This may be because of slight variation in W composition in the alloy in these areas. The two different areas are marked corresponding to regions with more 1D and 2D like modulation [Fig. 5 (c)]. The modulation vectors and lattice parameter for the supercell are indicated. The mechanism for 1D to 2D transition is shown in Fig. S3 (supplementary).

Density functional theory based calculation was carried out as a function of Mo and W concentration for $MoS_2$-$ReS_2$ and $WS_2$-$ReS_2$ alloy systems in both monolayer and bulk form. The data of cohesive energy against alloy composition for monolayer for two different alloys are given in Fig. 7 (a)-(b) (see supplementary for the bulk). The cohesive energy corresponding to 2H and unmodulated 1T polytypes are indicated by different colors points. All the 1T forms of unmodulated alloys are metallic and expected to have imaginary phonon mode similar to 1T $MoS_2$ and renders them unstable and undergo 2D Peierls structural modulation [23]. Therefore, for some alloy composition e.g., 25, 50 and 75 at.%, modulated $1T_d$ structural calculation was carried out to investigate the relative stability with respect to its un-modulated and 2H counterparts. One can observe that at ~50 at.% alloy composition a structural cross over between 2H to $1T_d$ takes place and the modulated structure is the most stable beyond that composition (see the data for 75 %). Therefore, it can be expected that for this composition range the alloy would be stable in the modulated $1T_d$ form. Experimental results suggest that for $MoS_2$-$ReS_2$ alloy system, the $1T_d$ modulated structure is stable even at 75 at.% in addition to 50 at.%, however, with a trimer type modulation. The initial calculations do not suggest the formation of such structural modulations and defects might be playing a role is stabilizing such structure through electronic charging effect [42]. It will be worth investigating both theoretically and experimentally the electronic property of such alloy superstructure for stable in-plane dipole domain structure for thinnest ferroelectric devices and beyond the scope of the present discussion [23]. Similarly, it will also be worth investigating any novel properties arising from the type II or trapezoid type modulation for 50% $MoS_2$-$ReS_2$ alloy system as part of the future exploration. The relative energy difference between the 2H and $1T_d$ form of structural poly-types at 50% alloy composition is 90, 40 meV for monolayer $MoS_2$-$ReS_2$ and $WS_2$-$ReS_2$ alloys, respectively. The relative energy difference between 2H semiconductor and 1T metallic phase of $MoS_2$ and $WS_2$ are ~ 0.25 eV and 0.35 eV, respectively, which are one order of magnitude higher than the 50% alloy case. The relative stability of 1T phase in the $SnS_2$-$MoS_2$ alloy system investigated earlier reported a cross over at ~60 at.% of Sn in $MoS_2$ lattice and all the alloys are found to be metallic in nature [31]. The modulated $1T_d$ structure corresponding to 50% alloy composition for both the cases is shown in Fig. 6 (a). The lattice parameters are indicated with the help of rhombus geometry defined earlier. The structure is semiconducting with a band gap of 0.317 eV and 0.115 eV eV for $MoS_2$-$ReS_2$ and $WS_2$-$ReS_2$ alloys, respectively [Fig. 7(b) & (c)].

However, with slightly different atomic configurations for the same composition, both metallic and semiconducting states are also found to be stable with smaller energy difference (supplementary). The difference between these later structural states and the most stable structural configuration is the atomic distances between two different cations. The most stable configuration is obtained if the Mo-Mo (or W-W) and Re-Re distances are different than the single value and this is because of the different sizes of the atoms which are also reflected in different lattice parameters of the individual crystals. The energy difference between semiconducting and the metallic ground states are very small, 90 and 40 meV for $MoS_2$-$ReS_2$ and $WS_2$-$ReS_2$ alloys, respectively. This is extremely important in the sense that by small external stimuli e.g., strain, charge or photon energy can tune the conductivity of such system with the slight modification of atomic arrangement within the same host lattice.

This can be useful as a switch, sensors, and many novel nanoscale electronic devices through metal-semiconductor transition [43, 44]. Though similar possibility exists between structural polytypes i.e. between 2H and 1T/1T$_d$ in the family, however, the energy difference is significantly higher i.e. ~ 0.25 to 0.35 eV and only alkali metal intercalation route has been shown to be able to control such structural transitions in a reversible way. This is in contrast to irreversible transitions in SnS$_2$-MoS$_2$ based alloy reported earlier where the similar transition is composition dependent [31]. The reversible metal-semiconducting transitions between H and T´ phases of W$_{0.67}$Mo$_{0.33}$Te$_2$ by charge mediation was also predicted theoretically with the small energy difference between the two structural polytypes [43].

The nature of valence and conduction band for 50% MoS$_2$/WS$_2$-ReS$_2$ alloy cases are briefly described below. The CBM and the VBM of MoS$_2$ are made of Mo $d_z^2$ orbital. In the case of ReS$_2$, both VBM and CBM have significant contributions from $d_{x^2-y^2}$ and $d_{xy}$ orbital. In case of MoS$_2$, the direct transition occurs from K (VBM) to K (CBM) and for ReS$_2$, it is from Γ (VBM) to Γ (CBM). For the two alloy cases, we have found stable semiconductor state for both Re$_{0.5}$Mo$_{0.5}$S$_2$ and Re$_{0.5}$W$_{0.5}$S$_2$ with the triclinic state having $P\bar{1}$ crystallographic symmetry. The VBM energy value at M point is comparable with R point but the CBM energy at M point is lowest for Re$_{0.5}$Mo$_{0.5}$S$_2$. Re$_{0.5}$Mo$_{0.5}$S$_2$ structure shows fundamental and direct band gap of 0.317 eV with fundamental transition between M to M [Fig. 7(b)]. An indirect transition from M to X having band gap of 0.115 eV is found for Re$_{0.5}$W$_{0.5}$S$_2$ [Fig. 7(e)]. Fig. 7(c) & (f) shows that for Re$_{0.5}$Mo$_{0.5}$S$_2$, the fundamental direct band transition has significant contribution from Mo $d_z^2$, and $d_{x^2-y^2}$ orbitals at VBM and $d_{xy}$ orbital at CBM and similarly the fundamental indirect band transition for Re$_{0.5}$W$_{0.5}$S$_2$, is primarily due to Re $d_{x^2-y^2}$ orbital at VBM and W $d_{xy}$ orbital at CBM. The band structure and DOS for other stable configurations are shown in the Fig. S7 and the related lattice parameters are listed in Table 2 (Supplementary).

## V. Conclusions

In summary, alloy formation, 2H vs. $1T_d$ structural stability and electronic structure between $MoS_2/WS_2$ and $ReS_2$ have been explored both by theoretically and experimentally. Theory indicates a structural cross over between 2H and $1T_d$ at approximately 50% alloy composition for both the system. Experimentally, two different structural modulations for 50% and a trimerize modulation is observed $MoS_2$-$ReS_2$ and only tetramerize modulation is observed for $WS_2$-$ReS_2$ alloy systems. The trimer modulation could offer the stable thinnest platform to explore the ferroelectric device. The 50% alloy system is found to be a suitable monolithic candidate for metal semiconductor transition with minute external perturbation. The results show yet another way to tune the various structural types and corresponding electronic structure in this important class of materials system.

**Acknowledgement**

The authors at JNCASR are grateful to Prof. CNR Rao for the constant support and advanced microscopy facility. R. Datta sincerely thanks KAUST aberration corrected microscopy core lab facility for Z-contrast imaging and a Sabbatical Funding for the visit. For computer time, this research used the resources of the Supercomputing Laboratory at KAUST (project k1143). S. S. thanks the Council of Scientific and Industrial Research for research fellowship and S. C. P. thanks the DST fast track (Grant SB/FT/Cs-07/2011) for the financial support.

**References**

[1] G. Eda, H. Yamaguchi, D. Voiry,T. Fujita, M. Chen, M. Chhowalla, Nano Lett. **11**, 5111−5116 (2012).

[2] T. Cao, G. Wang, W. Han, H. Ye, C. Zhu, J Shi, Q. Niu, P. Tan, E. Wang, B. Liu, J. Feng, Nat. Commun. **3,** 887−892 (2012).

[3] K. Mak, K. He, J. Shan, T. F. Heinz, Nat. Nanotechnol. **7,** 494−498 (2012).

[4] K. S. Novoselov , D. Jiang , T. Booth , V.V. Khotkevich , S. M. Morozov, A. K. Geim Proc. Natl. Acad. Sci **102,** 10415-10453 (2005).

[5] K. F. Mak, C. Lee, J. Hone, J. Shan, and T. F. Heinz, Phys. Rev. Lett. **102**, 136805 (2010).

[6] U. Maitra, U. Gupta, M. De, Ranjan Datta, A. Govindaraj, C. N. R. Rao, Angew. Chem. Int. Ed, **52,** 13057 – 13061 (2013).

[7] F. Wen, B. Li, Q. Wang, Y. Huang, Y. Gong, Y. He, P. Dong, J. Bellah, A. George, L. Ge, J. Lou, N. J. Halas, R. Vajtai, and P. M. Ajayan Nano Lett. **15 (1),** 259-265 (2015).

[8] M. Acerce, D. Voiry and M. Chhowalla Nat. Nanotech . **10,** 313-318(2015).

[9] H. S. S. R. Matte, A. Gomathi, A. K. Manna, D. J. Late, R. Datta, S. K. Pati, and C. N. R. Rao, Angew. Chem., Int. Ed. **49,** 4059, (2010).

[10] R. Sahu, D. Radhakrishnan, B. Vishal, D. S. Negi, A.Sil, C. Narayana, and R. Datta, arXiv:1610.05852

[11] S. Tongay, H. Sahin, C. Ko, A. Luce, W. Fan, Kai Liu, Jian Zhou,Ying-Sheng Huang, Ching-Hwa Ho, Jinyuan Yan, D. Frank Ogletree, Shaul Aloni, Jie Ji, Shushen Li, Jingbo Li, F.M. Peeters & Junqiao Wu, Nat. Commun. **5,** 3252, (2014).

[12] X. Fan, Y. Yang, P. Xiao, and W. Lau . J. Mater. Chem. A, **2,** 20545 (2014).

[13] T. Hu, R. Li, and J. Dong, J. Chem. Phys. **139,** 174702 (2013).

[14] K. K. Amara, Y. Chen, Y. Lin, R. Kumar, E. Okunishi, K. Suenaga, S. Quek, and G. Eda Chem. Mater **28,** 2308-2314 (2016).

[15] X. Fan, Y. Yang, P. Xiao, and W. Lau, J. Mater. Chem. A, **2,** 20545 (2014).

[16] A. Andersen, S. M. Kathmann, M. A. Lilga, K. O. Albrecht, R. T. Hallen, and D. Mei, J. Phys. Chem. C **116,** 1826–1832 (2012).

[17] X. Rocquefelte, F. Boucher, P. Gressier, G. Ouvrard, P. Blaha and K. Schwarz Phys. Rev.B **62,** 2397 (2000).

[18] X. Sun, Z. Wang, Z. Li, & Y. Q. Fu, Sci. Report **6,** 26666 (2016).

[19] L. Wang, Z. Xu, W. Wang, and X. Bai, J. Am. Chem. Soc. **136,** 6693−6697 (2014).

[20]. D. Yang, S. Jiménez Sandoval, W. M. R. Divigalpitiya, J. C. Irwin, and R. F. Frindt, Phys. Rev. B **43,** 12053 (1991).

[21] F. Wypych, Th. Weber, and R. Prins, *Chem. Mater.*, **10 (3),** 723–727 (1998).

[22]X. Rocquefelte, F. Boucher, P. Gressier, G. Ouvrard, P. Blaha, and K. Schwarz, Phys. Rev. B **62,** 2397 ( 2000).

[23] N. S. Shirodkar, and U. V. Waghmare, PRL **112,** 157601 (2014).

[24 B. Radisavljevic1, A. Radenovic, J. Brivio1, V. Giacometti, and A. Ki, Nat. Nanotech **6,** 147-150 (2011).

[25] Y. Rong, Y. Sheng, M. Pacios, X. Wang, Z. He, H. Bhaskaran, J. H. Warner, ACS Nano, **10,** 1093-1100 (2015).

[26] T. Park, J. Suh, D. Seo, J. Park, D. Lin, Y. Huang, H. Choi, J. Wu, C. Jang, and J. Chang Appl. Phys. Lett. **107,** 223107 (2015).

[27] M. Tahir, P. Vasilopoulos, and F. M. Peeters, Phys. Rev. B **93,** 035406, (2016).

[28] F. Wypych, Th. Weber, and R. Prins, Chem. Mater. **10,** 723-727 (1998).

[29] M. Kertesz, and Roald Hoffmann , J. Am. Chem. SOC. **106,** 3453-3460 (1984).

[30]. K. Dileep, R. Sahu, Sumanta Sarkar, Sebastian C. Peter, and R. Datta, J. Appl. Phys. **119,** 114309 (2016).

[31] F. Raffone, C. Ataca, J. C. Grossman,and G. Cicero, J. Phys. Chem. Lett. **7,** 2304−2309 (2016).

[32] L. Yadgarov, D. G. Stroppa, R. Rosentsveig, R. Ron, A. N. Enyashin, L. Houben, and R. Tenne Z. Anorg. Allg. Chem. **638 (15),** 2610–2616 (2012).

[33] A. A. Tedstone, D. J. Lewis, and P. O'Brien, Chem. Mater. **28,** 1965−1974 (2016).

[34] Qi -C. Sun, L. Yadgarov, R. Rosentsveig, G. Seifert, R. Tenne, and J. L. Musfeldt ACS nano **7,** 3506–3511 (2013).

[35] M. Chhetri, U. Gupta, L. Yadgarov, R. Rosentsveig, R. Tennec, and C. N. R. Rao Dalton Trans., **44,** 16399 (2015).

[36] K. Dolui, I. Rungger, C. D. Pemmaraju, and S. Sanvito, Phys. Rev. B **88**, 075420 (2013).

[37] A. N. Enyashin, L. Yadgarov, L. Houben, I. Popov, M. Weidenbach, R. Tenne, M. Bar-Sadan, and G. Seifer J. Phys. Chem. C **115,** 24586–24591 (2011).

[38] P. Blaha, K. Schwarz, G. K. H. Madsen, D. Kvasnicka, and J. Luitz, WIEN2k:An Augmented Plane Wave þ Local Orbitals Program for Calculating CrystalProperties (Karlheinz Schwarz, Techn. Universität Wien, Austria, 2001).

[39] S. Grimme, J. Antony, S. Ehrlich, and H. Krieg, J. Chem. Phys. **132,** 154104 (2010).

[40] J. P. Perdew, K. Burke, and M. Ernzerhof, Phys. Rev. Lett. **77**, 3865 (1996).

[41] Sang Wook Han, Won Seok Yun, J. D. Lee, Y. H. Hwang, J. Baik, H. J. Shin, Wang G. Lee, Young S. Park, and Kwang S. Kim, Phys.Rev B **92,** 241303, (R) (2015).

[42] Y. Lin, D. O. Dumcenco, Y. Huang, and K. Suenaga, Nat. Nanotech **9,** 391-396, (2014).

[43] C. Zhang, S. KC, Y. Nie, C. Liang, W. G. Vandenberghe, R. C. Longo, Y. Zheng, F. Kong, S. Hong, R. M. Wallace, and K. Cho, ACS Nano **10 (8),** 7370–7375 (2016).

[44] S. S. Chou, Y. Huang, J. Kim, B. Kaehr, B. M. Foley, P. Lu, C. Dykstra, P. E. Hopkins, C. J. Brinker, J. Huang, and V. P. Dravid, J. Am. Chem. Soc. **137,** 1742−1745 (2015).

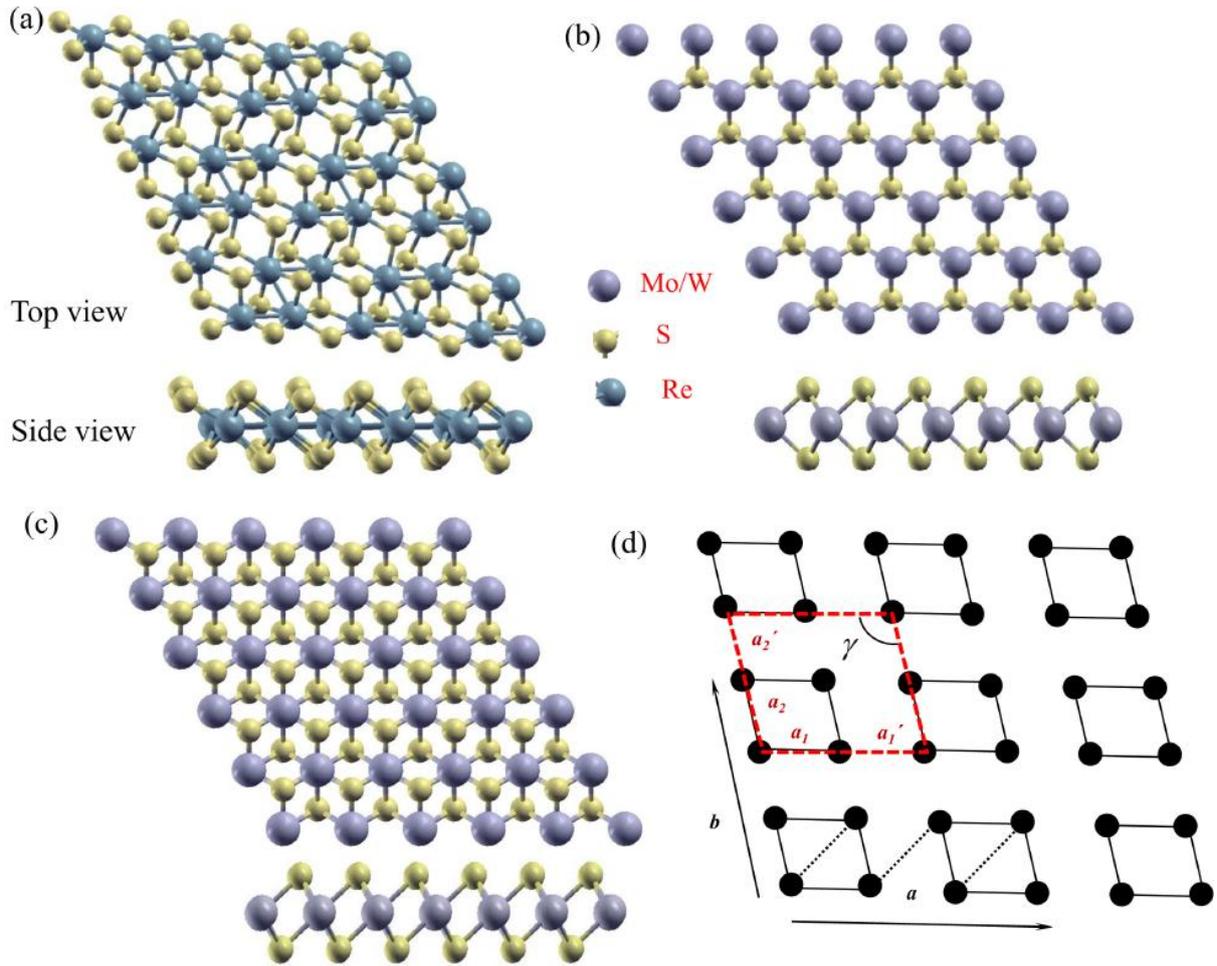

FIG. 1. Example schematic structural models of (b) 2H MoS$_2$/WS$_2$, (c) 1T MoS$_2$/WS$_2$, and (a) 1T$_d$ ReS$_2$ polytypes. The side view of each structural of polytype is also shown. (d) schematic supercell unit for the superstructure phases of 1T$_d$ structure with various structural parameters defined by $a_1$, $a_2$, $a_1'$, $a_2'$, $a$, $b$, and $\gamma$. These structural parameters are used throughout the text to define various superstructure are observed. Also, in this schematic is shown with various colors different types of metal-metal hybridization bonds possible in the systems.

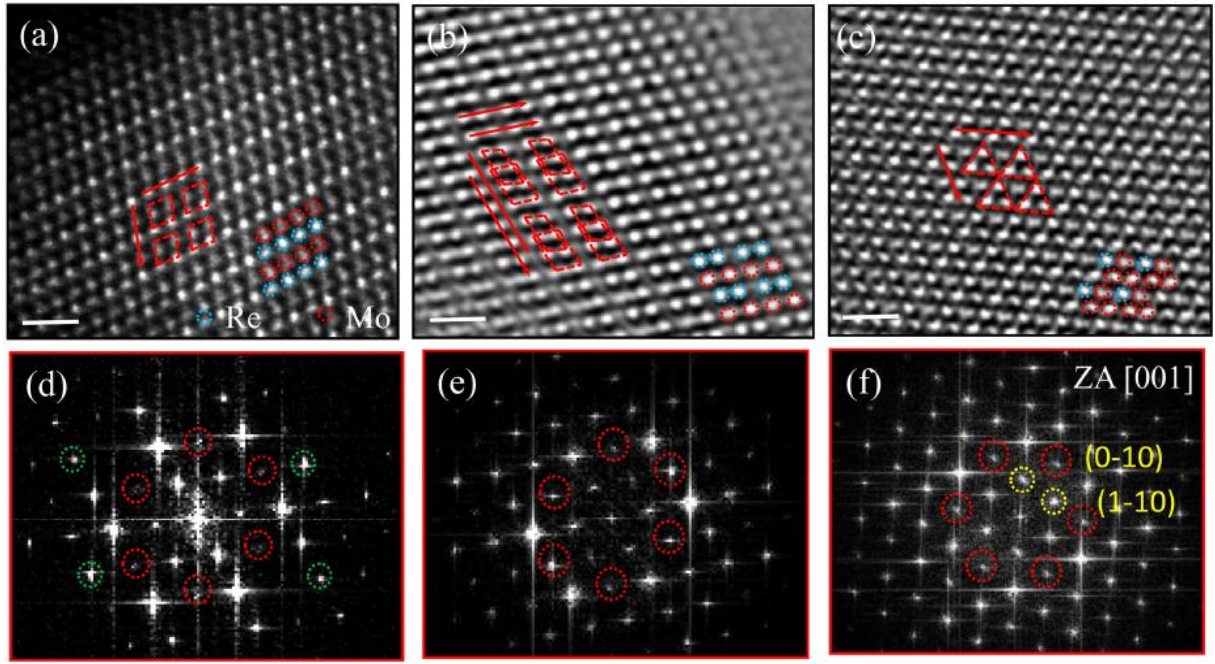

FIG. 2. Various types of superstructures observed in the case of MoS$_2$-ReS$_2$ alloy. Two different types of structural modulations are observed in 1T$_d$ phase with X$_{Re}$= 50%, (a) type I, where alternate rows are filled either by Re or Mo atoms giving rise to a superstructure which can be defined by 2*a*×2*a* supercell and (b) type II where the superstructure can be defined by two different 4*a*×2*a* supercells. The corresponding FFTs are shown in (d) and (e) where characteristic spots confirming the 1T phase and structural modulation are marked with red and yellow circles, respectively. The major difference between superstructuer (a) & (b) is two different metal-metal hybridization distance in the structure (see text for details). (c) A trimer structure modulation is also observed with X$_{Re}$= 25% areas where trimer made of Mo atoms forming corner sharing triangular arrengment. The corresponding FFT is shown in (f). The structureal parameters derived from experimental images are given in the supplementary and indicates they are more close to ideal hexagonal sytem in case of (c) and triclinic type in case of (a) & (b).The green circle refers super lattice spot. The scale bar is 0.5 nm.

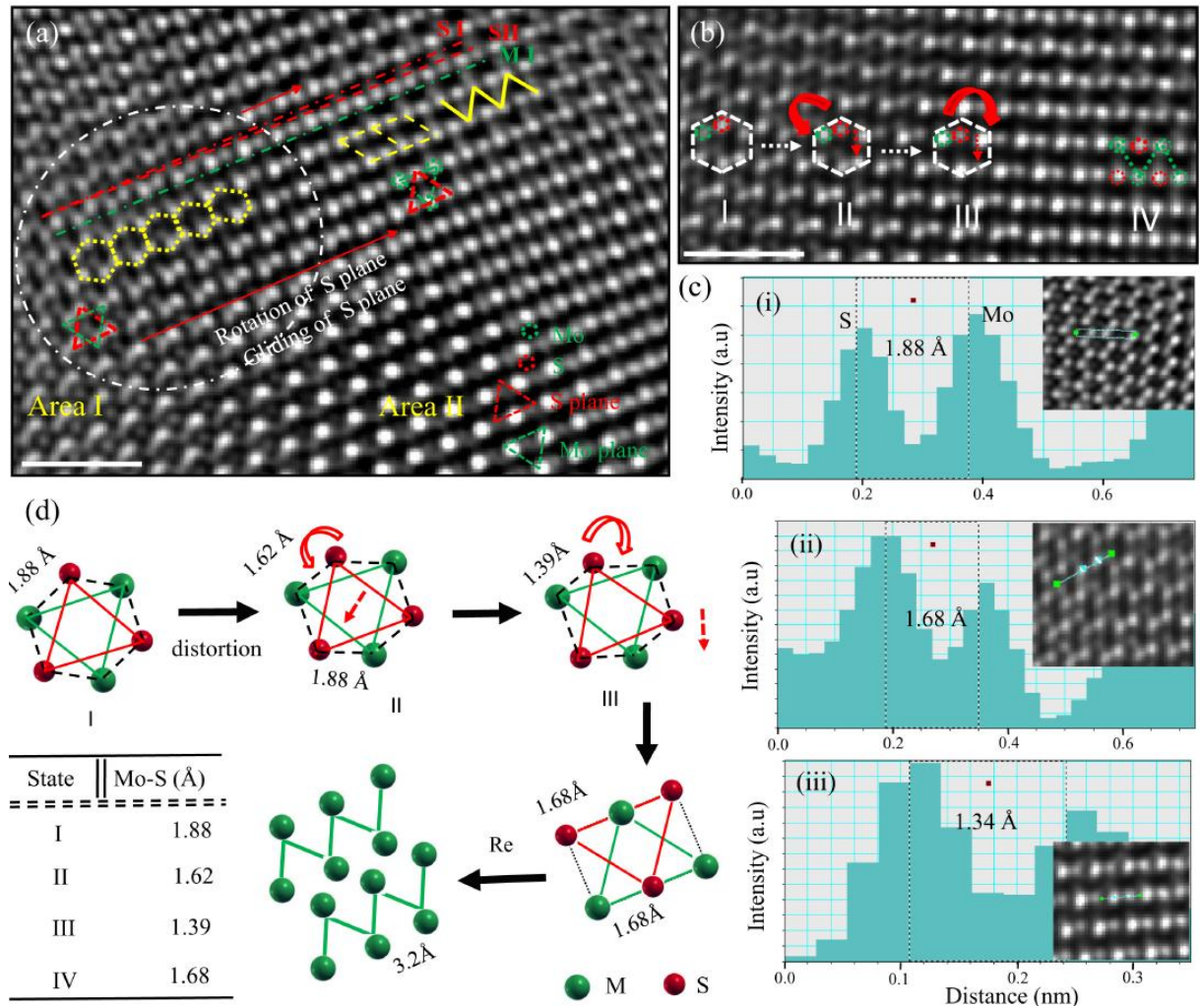

FIG.3. Figure (a) is the HRTEM image showing the pathway of phase transition from 2H to 1T$_d$ through intermediate metastable state. The phase transition occurs via rotation and gliding of upper S plane and is also indicated. SI and SII lines are the position of upper S plane before and after the S plane movement while the bottom S plane is static. Re doping makes cation hexagon to modulate as shown in yellow color. Figure (b) display the magnified view of gliding and rotation of upper S plane associated with intermediated phases. Figure (c) shows Mo-S bond length in distorted and undistorted MoS$_2$. Figure (d) is the schematic to represent evolution of 1T$_d$ phase from 2H through different intermediate phases. The experimentally derived metal-S bond lengths in different states are listed in the table. The scale bar is 1 nm.

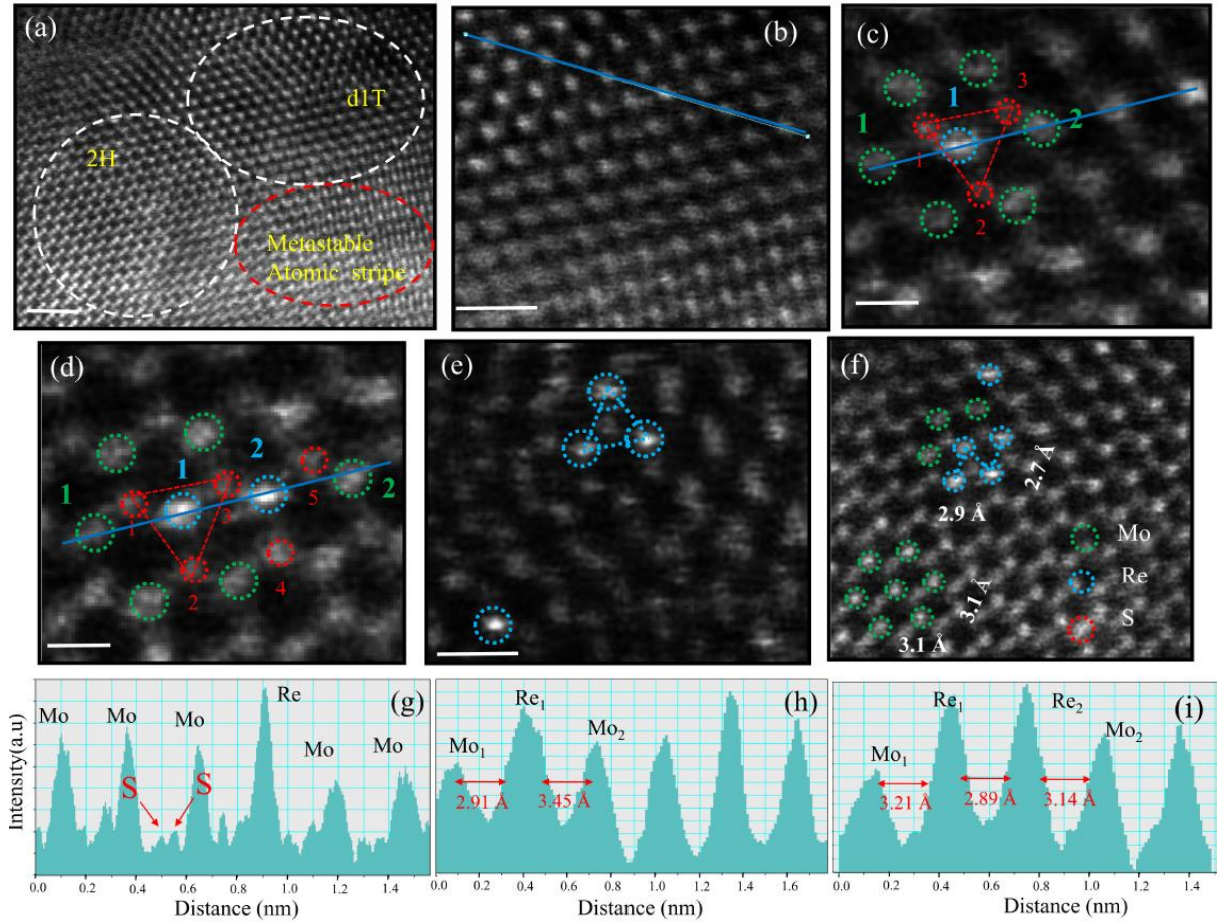

FIG. 4. (a) The STEM HAADF images show the phase transition from 2H to 1T$_d$ through intermediate metastable phases. This is equivalent to Figure. 3 but with the direct detection of Re atoms in the lattice. Figure (b) is the 1T$_d$ phase and the (g) corresponding line scan showing two S peaks in between two metal atoms. Figure (c)-(f) show monoatomic, two, three and four (tetramer geometry, Re$_4$) atom clusters in Re doped MoS$_2$. The figure (g) and (h) are the line scan showing Re dopant corresponding to figure (c) and (d). The scale bars are 0.5 nm for (a), and 0.2 nm for (b)-(f).

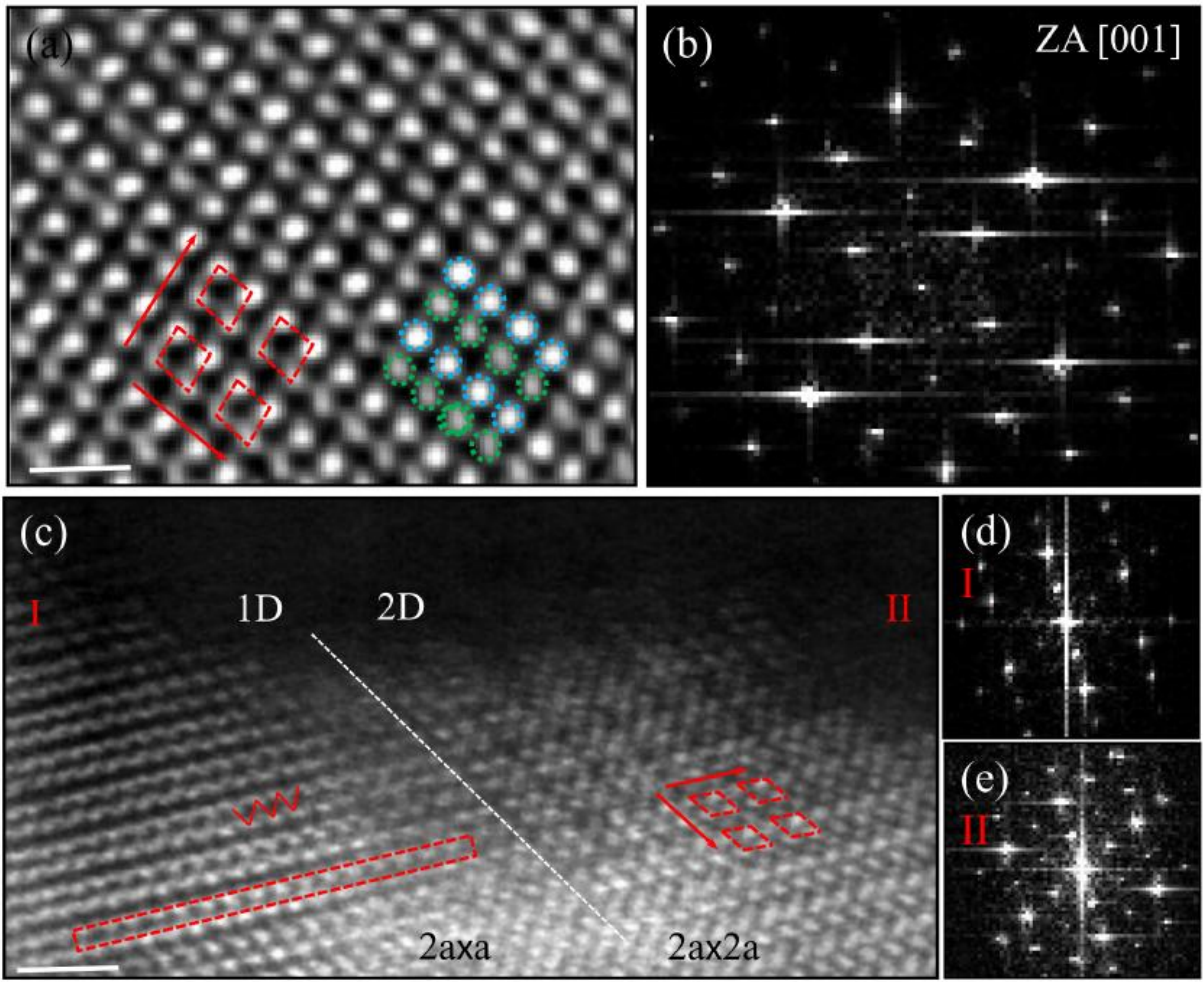

FIG. 5. (a) HRTEM image and (b) corresponding FFT showing $1T_d$ phase with structural modulation for $X_{Re}=$ 50% $WS_2$-$ReS_2$ alloy. This is similar to type I modulation observed in case of $MoS_2$-$ReS_2$ alloy. (c) Some areas shows strength of structural modulation along different directions i.e. 2D vs. 1D, probably due to slightly variation of W/Re concentration and the corresponding FFTs are shown in the (d) & (e). The scale bar is 1 nm.

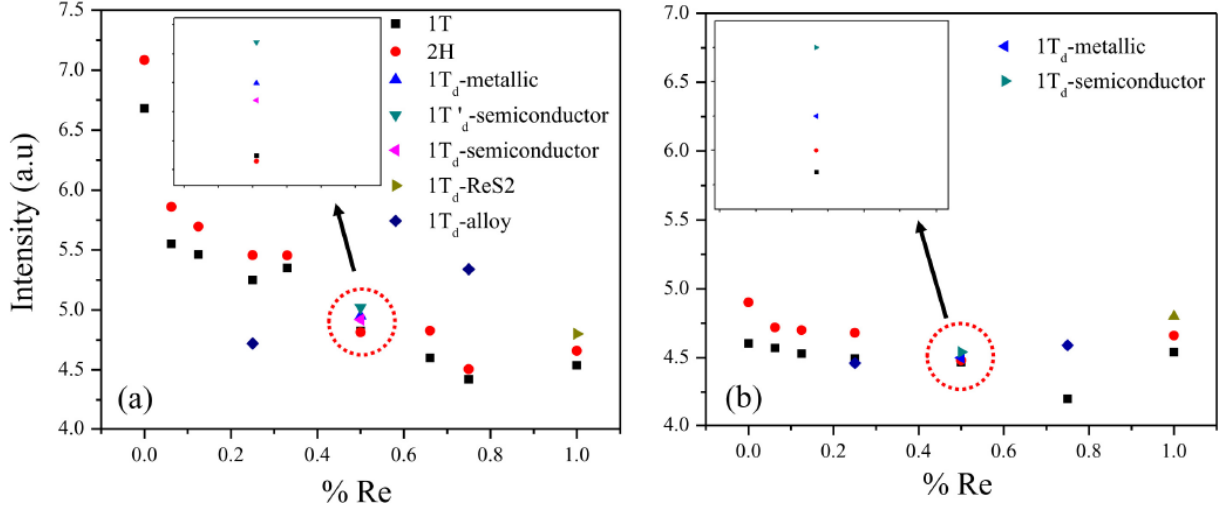

FIG. 6. Relative stability of 2H vs. 1T polytype of monolayer (a) $MoS_2$-$ReS_2$, (b) $WS_2$-$ReS_2$ alloys as a function of Re concentration. Structural modulation are considered some composition and the corresponding energy values are marked with different colors. Approximately, at $X_{Re}$ = 50%, a structural cross over is observed between 2H & 1T and 2H & $1T_d$ polytypes. Also, in (a) shows the energy values of different structural ground states for 50% $MoS_2$-$ReS_2$ alloys.

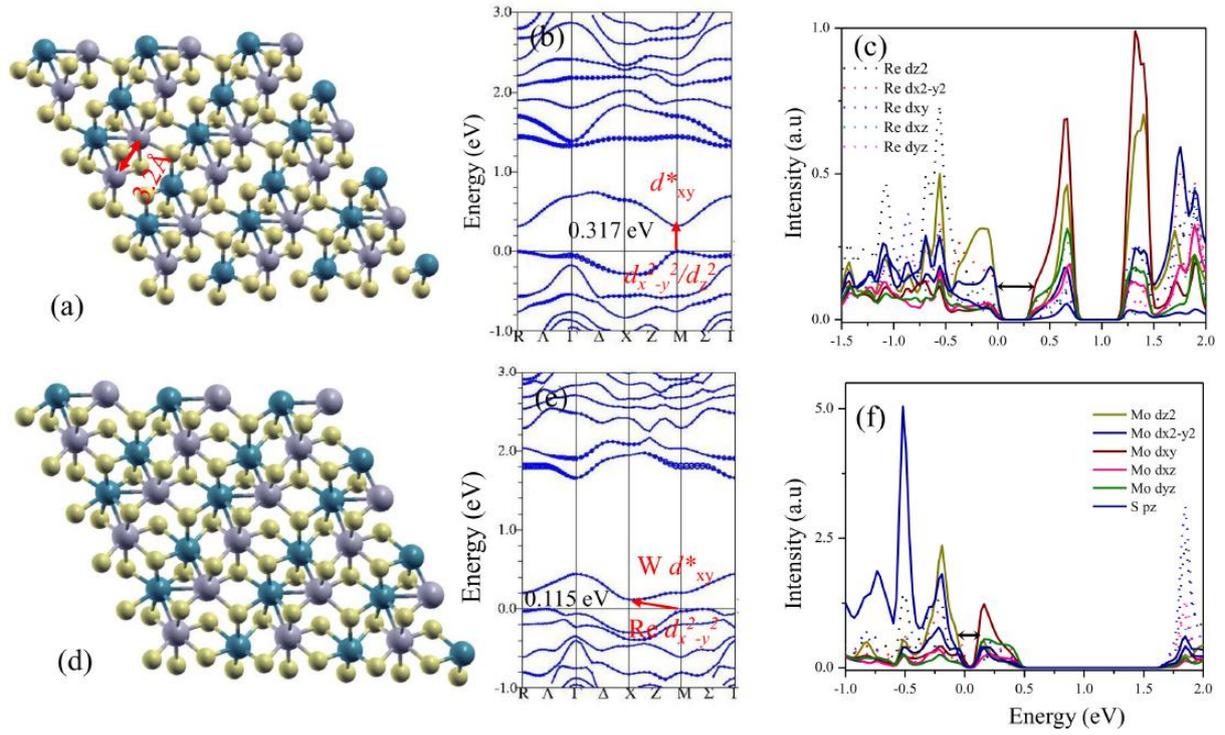

FIG. 7. The most stable $1T_d$ structure of $X_{Re}=50\%$ (a) $MoS_2$-$ReS_2$, (b) $WS_2$-$ReS_2$ alloy. Both the structures are semiconducting and the corresponding band structure and density of state are given (b) & (c), (e) & (f) for $MoS_2$-$ReS_2$, $WS_2$-$ReS_2$ alloys, respectively.

# Supplementary Information

Nature of low dimensional structural modulations and relative phase stability in $MoS_2$/$WS_2$-$ReS_2$ transition metal dichalcogenides alloys

R. Sahu,[1] U. Bhat,[1] N. M. Batra,[3] S. Horta,[1] B. Vishal,[1] S. Sarkar,[2] S. Assa Aravindh,[3] S. C. Peter,[2] I. S. Roqan,[3] P. M. F. J. D. Costa,[3] and R. Datta[1,3,*]

[1]*International Centre for Materials Science, Chemistry and Physics of Materials Unit, Jawaharlal Nehru Centre for Advanced Scientific Research, Bangalore 560064, India.*

[2]*New Chemistry Unit, Jawaharlal Nehru Centre for Advanced Scientific Research, Bangalore 560064, India.*

[3]*Physical Science and Engineering Division, King Abdullah University of Science and Technology (KAUST), Thuwal, 23955-6900, Saudi Arabia.*

*Corresponding author E-mail: ranjan@jncasr.ac.in*

**I .Relative phase stability of Re doped bulk WS$_2$ and MoS$_2$:**

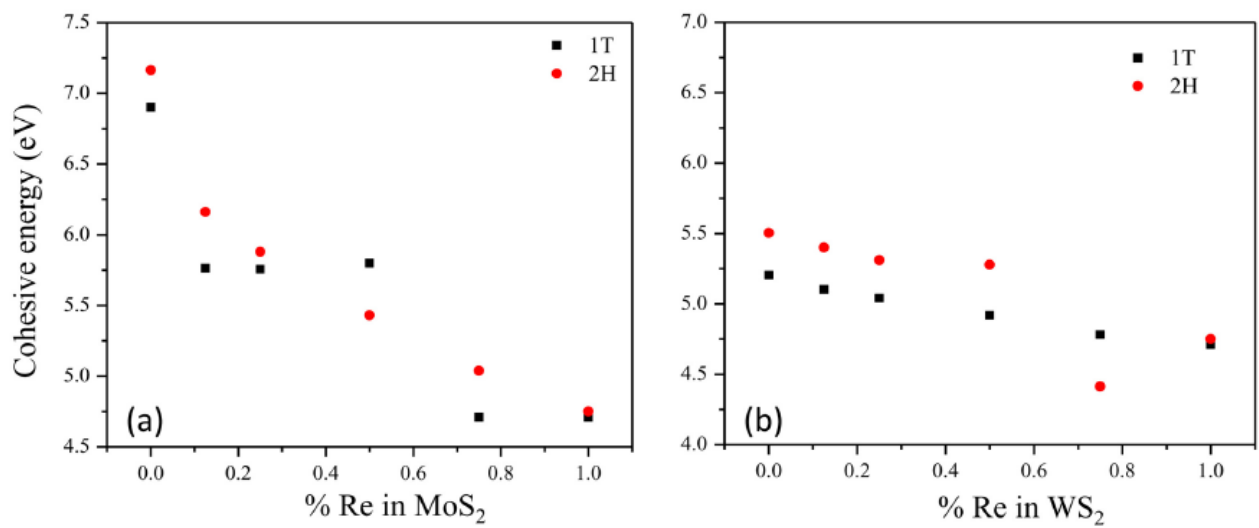

**Figure S1**. The diagram shows the relative stability of MoS$_2$-ReS$_2$ and WS$_2$-ReS$_2$ alloys with Re concentrations. No structural modulation is considered in this case. Kindly refer to the diagram corresponding to the monolayer for the same.

## II. Effect of Re doping on structural modulation, an experimental observation:

### HAADF Z-contrast imaging of $Re_{0.5}Mo_{0.5}S_2$ alloy with Type I modulation

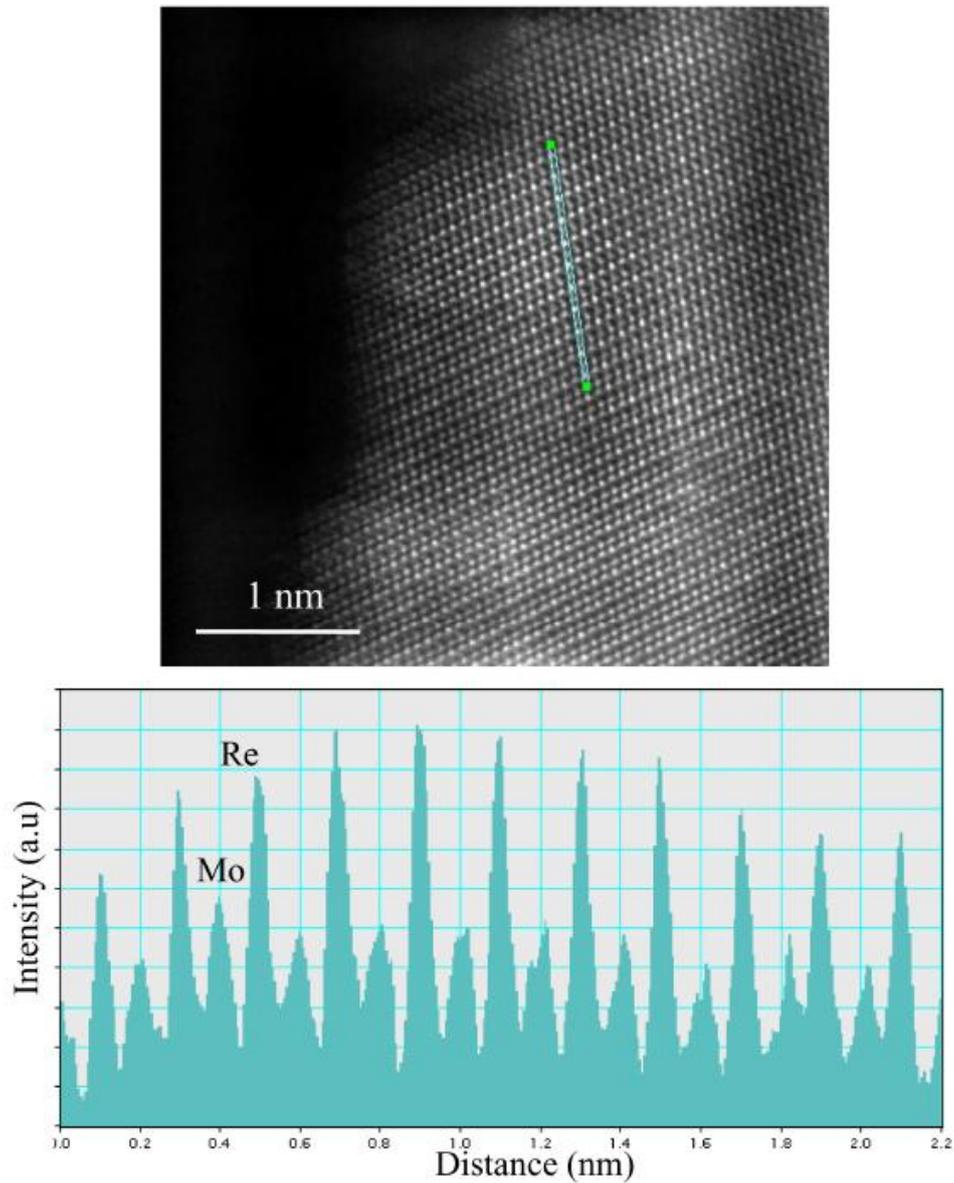

**Figure S2**. (a) Z-contrast HAADF images of 50% $MoS_2$-$ReS_2$ alloy with type I structural modulation. (b) Line profile along a row of atoms identifying Re and Mo atoms occupying alternately in the lattice.

**Structural modulation of different strength along two different directions or in other words relatively more 2D or 1D modulation:**

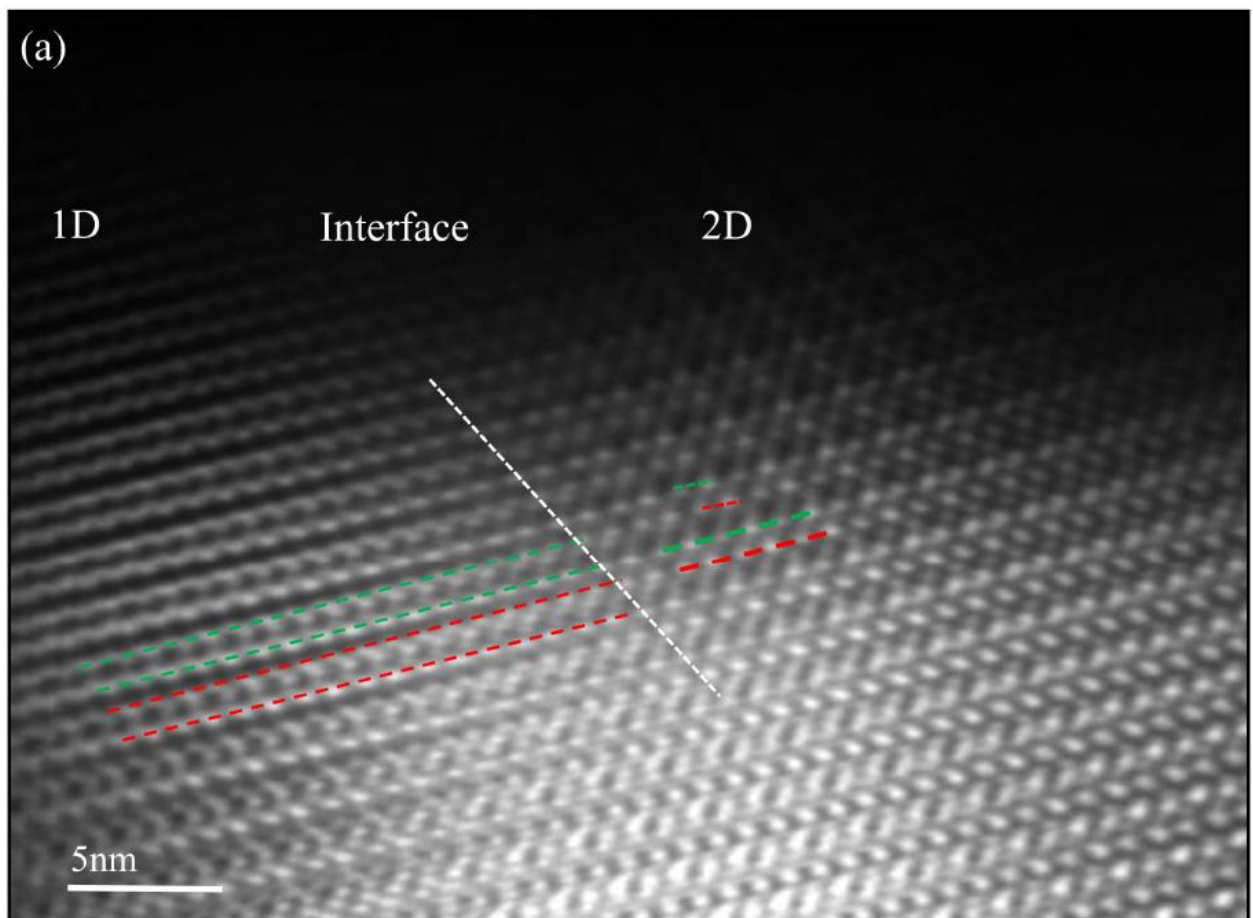

**Figure S3.I**. Fourier filtered Z-contrast HAADF image of 1T´´ (strong 2D chain or moduation) and 1T´ (more 1D chain or modulation) areas in Re doped WS$_2$. This may be because of slight variation in Re in WS$_2$ lattice.

From the above image one can observe clearly the two different structural modulations in terms of relative strength of modulation along two different directions or 1D and 2D chain structure. Two units of 1D chain is indicated by green and red parallel lines with spacing between each chain is 3.4 Å . The distance of atoms inside 1D chain keep increasing near the interface and recombine with another 1D chains and form 2D dimond chain like structure crossing the inter face. The existence of 1D-2D system is because of slight variation of Re in $WS_2$. From line profile we found that near the interface Re percentage is more and randomly distributed. This random distribution high concentration stabilized the system with strong modulation along two directions or 2D diamond chains.

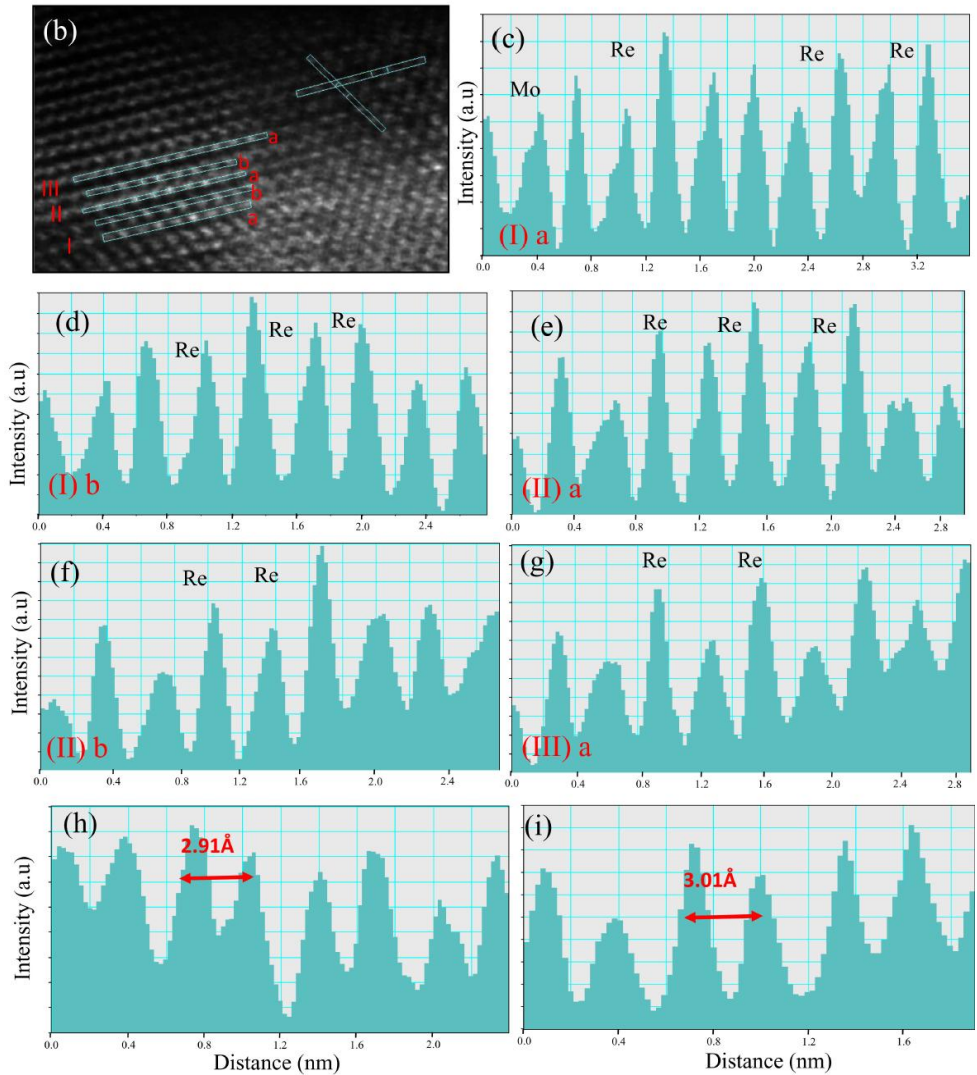

**Figure S3.II.** (b) is the same STEM HADDAF image Figure S3.I. (c)-(g) The line scan profiles are marked as I, II, III to show Re incorporation. (h) and (g) are the line scan measurement from the 2D area corresponding to two different directions. (c)-(g) shows the increase % of Re incorporation near the interface.

**Trimer phase in $Mo_{0.75}Re_{0.25}S_2$**

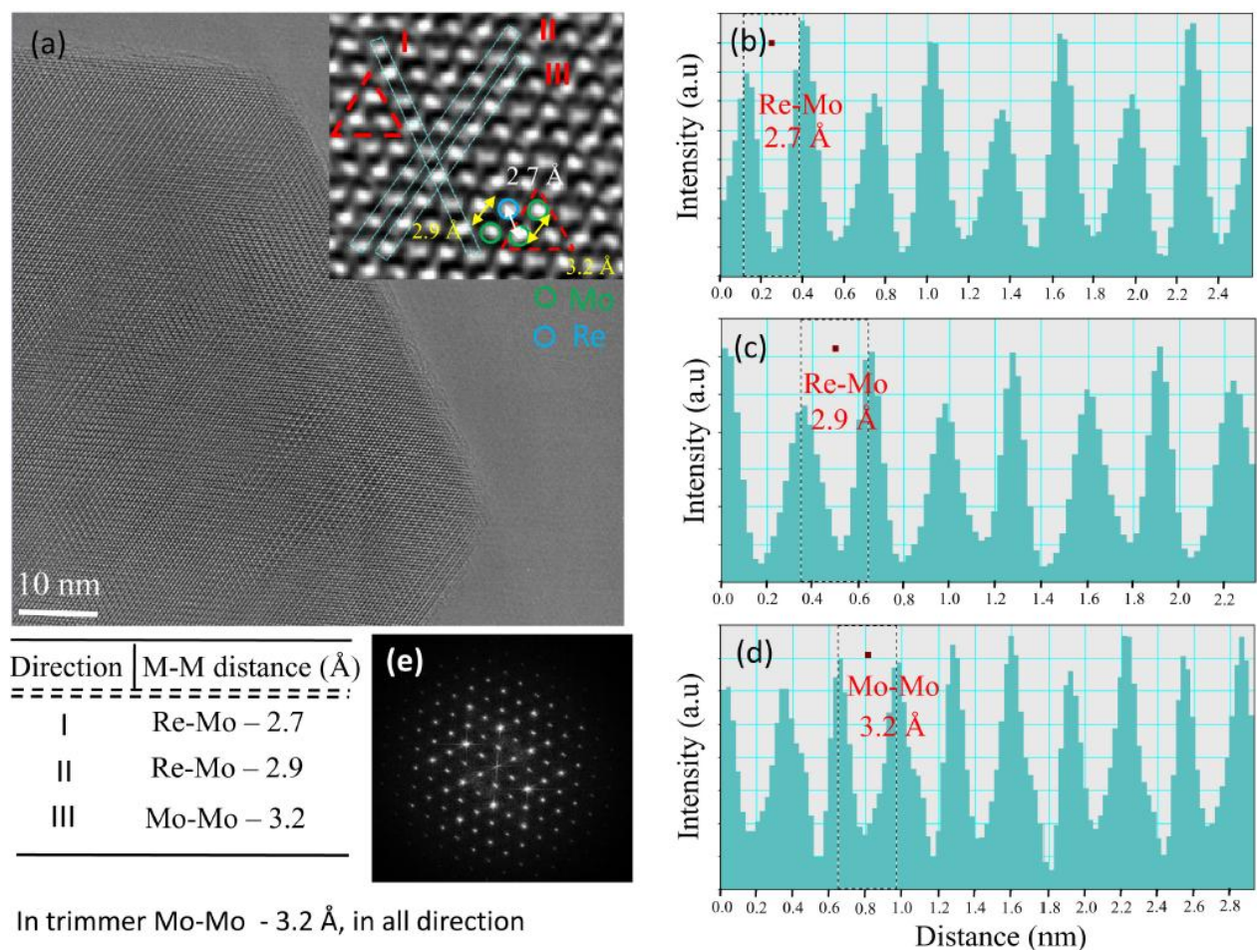

**Figure S4.** (a) Low magnification HRTEM image of Trimer. The red and green color indicate Re and Mo atoms respectively. (b) – (c) are line scan to show distances from metal-metal bond from different direction mentioned in the inset. (e) shows the FFT of figure (a). Metal-Metal distances are listed in the table

Theoretically the trimer phase of $MoS_2$ was predicted to be thinnest ferroelectric. The trimer phase is the most unstable states among various modulated $1T_d$ form of MoS2. From HRTEM images of $Mo_{0.75}Re_{0.25}S_2$ its clear that in Re doping induces Pierls distortion in 2H $MoS_2$ causing Mo atom to trimerize with equilateral geometry with Mo-Mo distance 3.2 Å.

**2H to distorted 1T via intermediate steps in $Mo_{0.5}Re_{0.5}S_2$ and Re incorporation :**

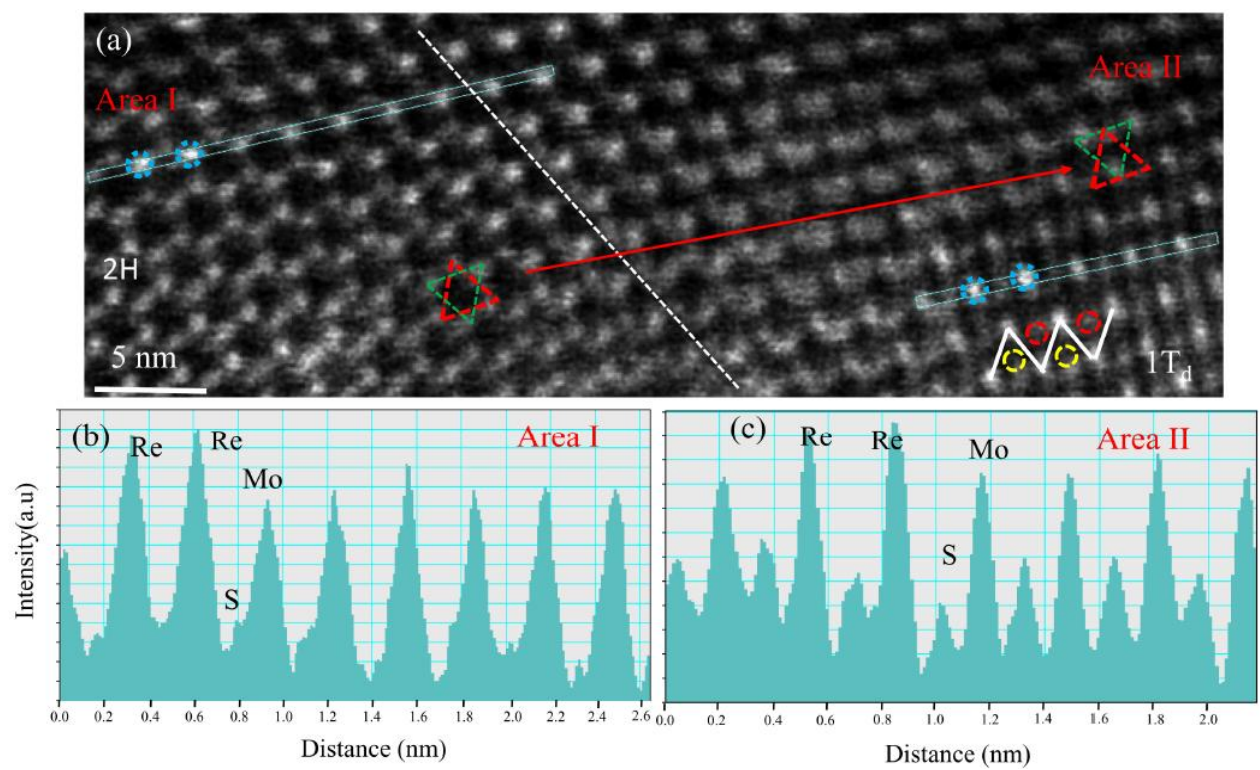

**Figure S5**. Figure (a) shows Re incorporation in 2H phase and in the intermediate phases. The line profile shows the atomic identification in the lattice.

In this Z contrast HAADF images, it is clearly shown further incorporation of Re atoms in intermediate phase give rise to $1T_d$. The Re incorporation in lattice creates lattice compression as well as lattice distortion. This distortion shows atomic rearrangement depending or Re concentration. Area I is 2H phase and the area II is the mixture of intermediate phase and $1T_d$ phase. The Z depending line profile from two areas confirm the Re incorporation in lattice.

**The schematic of 2H→1T$_d$ phase transition with and without metastable state in Re doped MoS$_2$ or WS$_2$:**

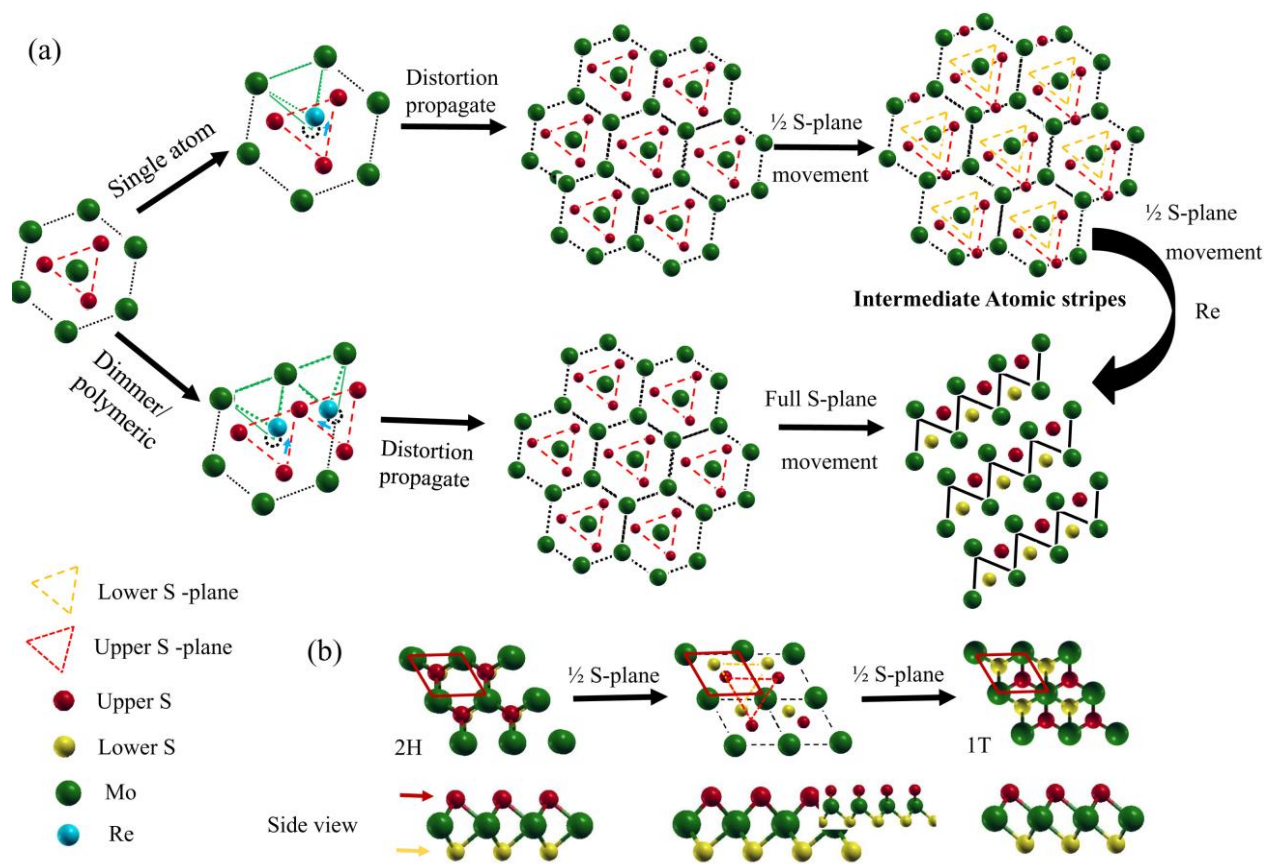

**Figure S6.** The Schemtic representations shows the route of polytype phase trasitions from 2H to 1T$_d$ through various intermediate phaes . (a) shows two possible route for transition depending on Re concentration, I) Monoatomic Re doping, and II) di or polyatom doping in 2H. (b) shows the side view of (a). The uper sulphur is displayed with red color.

The Re-Re and Re-metal (Mo/W) bond length being shorter then Mo-Mo / W-W bond length. The single Re incorporation of Re atoms where Re-Re bonding is absent, stabilize the metastable intermediate atomic stripe step due to ½ path movement of any S plane. The distortion due to Re incorporation in lattice propagates through the whole crystal. The blue arrow in the schematic shows the shifted position of Re atoms, which creates smaller Re-Metal bonds in distorted hexagon. The Re in corporation in the intermediate metastable shows phase transition to $1T_d$ due to another ½ path movement of S plane. In case of polyatomic doping, where Re-Re bonds are present, transforms to $1T_d$ phase directly through complete movement of S-plane. The displacement of S plane in three different phases is shown in figure (b). These lattice compression (distortion) due to Re is followed by atomic rearrangement through intermediate state and 2H to 1T phase transition occurs.

**Table 1. Various Metal-metal and metal-sulpher distances for different Re-Re clustering areas observed in figure 4 listed in this table (a) and (b).**

(a)

| Bond | M-S (Å) |
|---|---|
| Re-S$_1$ | 1.39 |
| Re-S$_2$ | 2.22 |
| Re-S$_3$ | 1.74 |
| Re-Mo$_1$ | 2.91 |
| Re-Mo$_2$ | 3.45 |

(b)

| Bond | M-S (Å) |
|---|---|
| Re$_1$-S$_1$ | 1.76 |
| Re$_1$-S$_2$ | 2.20 |
| Re$_1$-S$_3$ | 1.65 |
| Re$_2$-S$_3$ | 1.42 |
| Re$_2$-S$_4$ | 2.11 |
| Re$_2$-S$_5$ | 1.97 |
| Re$_1$-Re$_2$ | 2.89 |
| Re$_1$-Mo$_1$ | 3.21 |
| Re$_2$-Mo$_2$ | 3.14 |

## III) Variation in electronic properties due to different atomic arrangement in Re doped system:

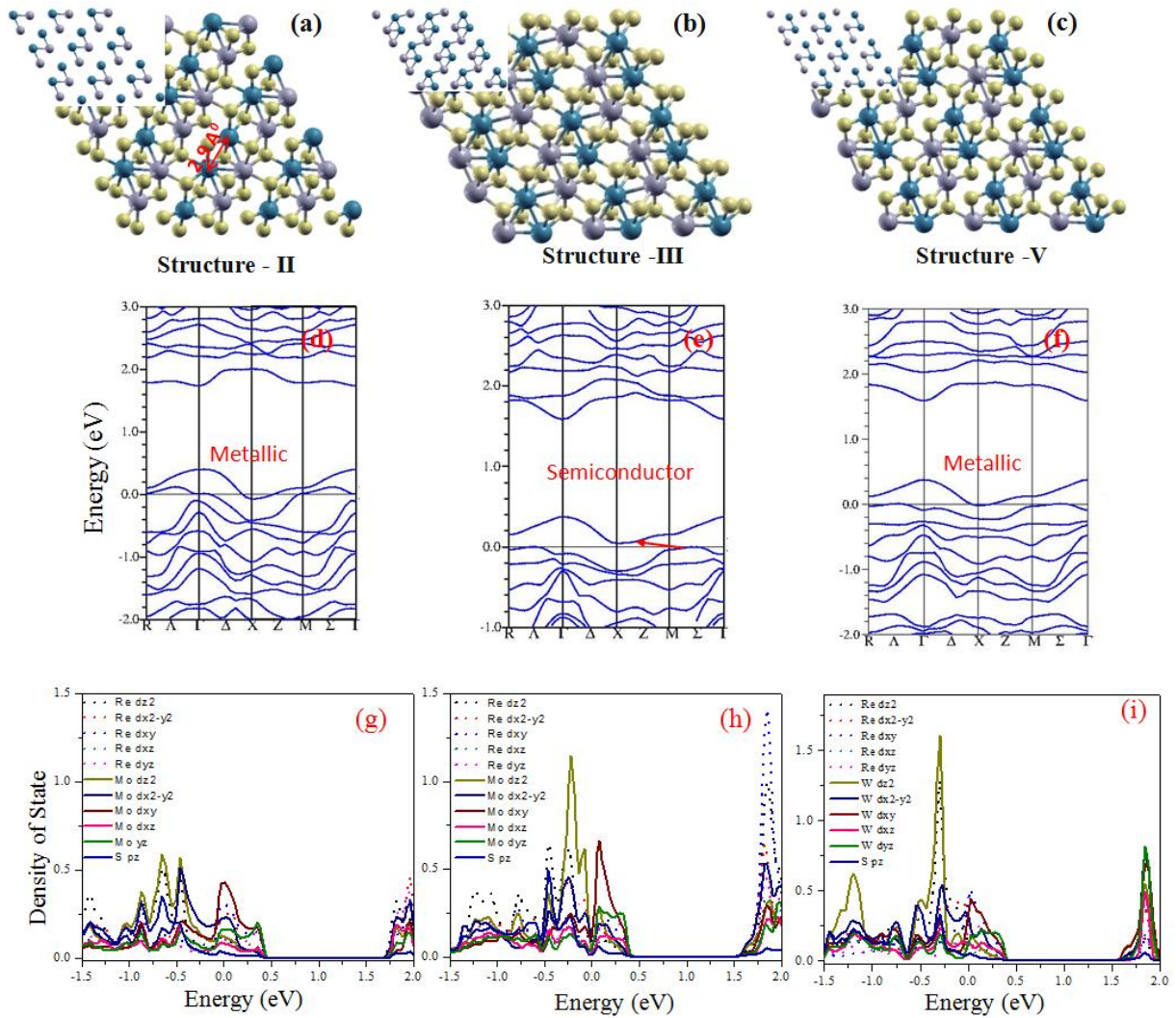

**Figure S7**. $2a \times a$ superstructures with slightly variation of structural parameters, (a) metallic $Mo_{0.5}Re_{0.5}S_2$ and (b) semiconducting ground states of $Mo_{0.5}Re_{0.5}S_2$. (c) $2a \times a$ superstructure with metallic $W_{0.5}Re_{0.5}S_2$. Band structure and DOS are given below corresponding structures. The relative energy differences between various structural configurations are given in the Table 2.

Figure (a) and (b) are the distorted modulated structure of $Re_{0.5}Mo_{0.5}S_2$. The cohesive energies of (a) is 30 meV higher than the semiconductor structure (b). So the metal to semiconductor transition is possible in room temperature. We started to find more stable

structure to stabilized semiconductor one, shown in Fig. 8. In this case the energy difference between the metallic state and semiconductor is 90 mev. The figure (f) shows the metallic nature having 40 meV less cohesive energy than semiconductor for $Re_{0.5}W_{0.5}S_2$. There is a possibility of another direct transition at Γ point above Fermi level in all those structure

**Table 2. The Lattice parameter and cohesive energy of simulated structures.**

| Structure | $a$ (Å) | $b$ (Å) | $\gamma°$ | Cohesive energy (eV) |
|---|---|---|---|---|
| Structure-I   | 6.167 | 6.224 | 119.03 | 4.94 |
| Structure-II  | 6.452 | 6.513 | 119.03 | 4.91 |
| Structure-III | 6.452 | 6.513 | 119.06 | 5.04 |
| Structure-IV  | 6.472 | 6.534 | 119.03 | 4.54 |
| Structure-V   | 6.484 | 6.543 | 119.03 | 4.50 |

Figure 7 (a) and (b) are structure-I and structure-IV respectively.

**Table 3. The experimental parameters calculated using Fig. 1(d) listed here.**

|  | $a_1$(Å) | $a'_1$(Å) | $a_2$ (Å) | $a'_2$ (Å) | $\gamma°$ |
|---|---|---|---|---|---|
| **Figure 2 (a)** | 2.90 | 2.98 | 2.68 | 3.18 | 74 |
| **Figure 2 (b)** | 2.41* | 3.22 | 5.14 |  | 114 |
|  | 2.7** | 2.98 | 5.14 |  |  |
| **Figure 2 (c)** | 2.78 | 3.5 | 3.2 | 3.33 | 109 |
| **Figure 4** | 2.42 | 2.71 | 2.43 | 2.71 | 80 |
| **Figure 5** | 2.68 |  | 3.1 | 3.47 | 109 |
|  | 2.77 | 3.53 | 2.82 | 3.71 |  |

- \* For Re atom
- \*\* For Mo atom